\documentclass[oldversion]{aa}  
\usepackage{graphicx}
\usepackage{natbib}
\setcitestyle{aysep={}} 
\usepackage{hyperref}
\usepackage{longtable}
\usepackage{lscape}
\begin{document}

   \title{Imaging of exocomets with infrared interferometry
}


   \author{Markus Janson\inst{1} \and
          Jayshil Patel\inst{1} \and
          Simon C. Ringqvist\inst{1} \and
          Cicero Lu\inst{2} \and
          Isabel Rebollido\inst{3} \and
          Tim Lichtenberg\inst{4,5} \and
          Alexis Brandeker\inst{1} \and
          Daniel Angerhausen\inst{6,7,8} \and
          Lena Noack\inst{9}
          }

   \institute{Department of Astronomy, Stockholm University, AlbaNova University Center, 10691 Stockholm, Sweden\\
              \email{markus.janson@astro.su.se}
        \and
        Department of Physics and Astronomy, Johns Hopkins University, 3400 N. Charles Str, Baltimore, MD 21218, USA
        \and
        Space Telescope Science Institute, 3700 San Martin Dr., Baltimore, MD 21218, USA
        \and
        Department of Physics, University of Oxford, Oxford OX1 3PU, UK
        \and
        Kapteyn Astronomical Institute, University of Groningen, PO Box 800, 9700 AV Groningen, NL
        \and
        ETH Zurich, Institute for Particle Physics \& Astrophysics, Wolfgang-Pauli-Str. 27, 8093 Zurich, Switzerland
        \and
        National Center of Competence in Research PlanetS (www.nccr-planets.ch)
        \and
        Blue Marble Space Institute of Science, Seattle, United States
        \and
        Freie Universit{\"a}t Berlin, Institute of Geological Sciences, Malteserstr. 74-100, 12249 Berlin, Germany
        }

   \date{Received ---; accepted ---}

   \abstract{
   Active comets have been detected in several exoplanetary systems, although so far only indirectly, when the dust or gas in the extended coma has transited in front of the stellar disk. The large optical surface and relatively high temperature of an active cometary coma also makes it suitable to study with direct imaging, but the angular separation is generally too small to be reachable with present-day facilities. However, future imaging facilities with the ability to detect terrestrial planets in the habitable zones of nearby systems will also be sensitive to exocomets in such systems. Here we examine several aspects of exocomet imaging, particularly in the context of the Large Interferometer for Exoplanets (\textit{LIFE}), which is a proposed space mission for infrared imaging and spectroscopy through nulling interferometry. We study what capabilities \textit{LIFE} would have for acquiring imaging and spectroscopy of exocomets, based on simulations of the \textit{LIFE} performance as well as statistical properties of exocomets that have recently been deduced from transit surveys. We find that for systems with extreme cometary activities such as $\beta$~Pictoris, sufficiently bright comets may be so abundant that they overcrowd the \textit{LIFE} inner field of view. More nearby and moderately active systems such as $\epsilon$~Eridani or Fomalhaut may turn out to be optimal targets. If the exocomets have strong silicate emission features, such as in comet Hale-Bopp, it may become possible to study the mineralogy of individual exocometary bodies. We also discuss the possibility of exocomets as false positives for planets, with recent deep imaging of $\alpha$~Centauri as one hypothetical example. Such contaminants could be common, primarily among young debris disk stars, but should be rare among the main sequence population. We discuss strategies to mitigate the risk of any such false positives.
   }

\keywords{Infrared: planetary systems -- 
             Comets: general -- 
             Planets and satellites: detection
               }

\titlerunning{Imaging of exocomets}
\authorrunning{M. Janson et al.}

   \maketitle
%

\section{Introduction}
\label{s:intro}

Rocky and icy bodies in the size range of $\sim$1 km to a few hundreds of kilometers are very abundant in our Solar System, and they are often concentrated in various locations such as the asteroid belt, the Kuiper belt, the Oort cloud, and the Lagrangian L4 and L5 points of the giant planets. They can be collectively referred to as ``planetesimals'', and are thought to be a natural outcome of the planet formation process \citep[e.g.,][]{schlichting2013,johansen2015}. Due to their small sizes and low temperatures, individual planetesimals are typically incredibly difficult to detect, even in the Solar System  (except in favorable circumstances), let alone in other stellar systems. However, planetesimals are occasionally disrupted, either through collisions with other planetesimals \citep[e.g.,][]{jackson2014}, or through tidal disruption during close encounters with planets \citep[e.g.,][]{cataldi2018,janson2020}. The dust produced as part of the disruption has a much larger collective optical surface than the planetesimal itself, and it can therefore produce a visible signal \citep[e.g.,][]{lawler2015}. The dust from an individual disruption disperses over a short timescale due to irradiation from the central star, but a rich planetesimal belt, where disruptions occur sufficiently frequently, can form a continuously visible structure known as a debris disk. Debris disks around other stars typically need to be substantially brighter than the Kuiper belt to be observable, but still, they are relatively common \citep[e.g.,][]{eiroa2013}. For example, $\sim$16--17\% of Sun-like stars host confirmed debris disks, with a mildly increasing trend toward higher-mass stars \citep{trilling2008,sibthorpe2018}. Additionally, stacking thousands of phase-folded light curves of Kepler planet candidate systems has revealed putative flux deficits around the L4 and L5 points of the planets, which would imply that populations of trojan planetesimals could be common in planetary systems \citep{hippke2015}. These observational results imply that a large number of planetesimals must be an essentially ubiquitous feature of stellar systems, as is expected in core accretion-based planet formation theories \citep{drazkowska2022}. Gaining observational insight into the compositional properties of planetesimals will constrain the chemical makeup of rocky exoplanets in inner planetary systems \citep{krijt2022}, as the timescale and physical mechanisms of planetesimal differentiation influence the atmospheric composition of mature exoplanets \citep{lichtenberg2022pp7}.

Through dynamical interactions, for example with planets in the same system \citep{duncan1997}, a fraction of the planetesimals can be scattered into highly eccentric orbits. If the periastron is close enough to the parent star (on the order of 1 au or closer), then for a fraction of the orbit, the planetesimal partially evaporates and forms a cloud of dust and gas around it -- that is to say, a coma, making it identifiable as a comet. During this phase, the effective optical surface of the object increases drastically due to the coma, and from the heating of the star, the temperature also rises, making the comet many orders of magnitude brighter in both visible and infrared light than in its ``normal'' quiescent phase.

Even though active comets are much brighter than planetesimals of corresponding sizes, they are still faint relative to the parent stars. Furthermore, in order to receive enough stellar flux to be reasonably bright at visible and infrared wavelengths, potentially observable comets need to reside at a small separation from the parent star, placing them in the same extreme contrast regime as directly imaged planets \citep[with the exception of young, self-luminous planets, e.g.][]{marois2010,janson2021a,bonati2019}. Indeed, as discussed in Sect. \ref{s:observability}, moderate-sized planets (approximately Earth- to Neptune-sized) overlap with the brightness range of large comets. Exocomets are therefore subject to the same type of high-contrast problem as exoplanets, where the vast majority of them are beyond reach for direct imaging with present-day observational facilities. Hence, for exocomets and exoplanets alike, the first pieces of evidence have come from indirect methods. In the case of exocomets, this evidence has arrived through transit measurements of comets (or material of cometary origin) passing in front of the stellar surface, as summarized in Sect. \ref{s:observability}. However, as technology approaches the capability to directly image terrestrial planets, it simultaneously approaches the capacity to directly image exocomets. 

The Large Interferometer for Exoplanets \citep[\textit{LIFE}, see][]{quanz2022} is a mission concept for a space-based infrared interferometer capable of nulling interferometry. The primary aim of the conceived mission is to image planets through infrared nulling interferometry, being optimized for sufficient contrasts to detect and characterize several tens of Earth-like planets in the classical habitable zones of their parent stars \citep[e.g.,][]{konrad2022,alei2022}. The resolution, sensitivity and contrast provided by such a mission would also open up a wide range of other scientific avenues that could not be addressed with any other existing or planned facility. In this paper, we study  observational aspects of exocomets in the context of high-contrast imaging surveys, with a strong focus on \textit{LIFE} since it is foreseen as being particularly potentially powerful for the detection of comets. There are two recent developments that make a concrete study to this end possible for the first time: Firstly, a \textit{LIFE} simulator has been produced and made available, such that realistic estimations can be performed for the mission performance and required integration times \citep[LIFEsim\footnote{\url{https://lifesim.readthedocs.io/}}; see][]{dannert2022}. Secondly, a large number of exocomet transits have now been identified in the $\beta$ Pictoris system \citep{zieba2019,lecavelier2022}, making it possible for the first time to predict the frequency and observational properties of comets in this young debris disk system.

The paper is outlined as follows: In Sect. \ref{s:mission}, we discuss the basic parameters assumed for \textit{LIFE} when used in an exocometary context. In Sect. \ref{s:observability}, we outline previous observational evidence for exocomets and what has been learned from them, and how this knowledge is used to predict observational properties for direct imaging purposes. A set of particularly promising potential targets for exocomet searches are described in Sect. \ref{s:targets}. We then present the results from a case study for the $\beta$ Pic system in Sect. \ref{s:casestudy}, which sets the stage for a wider discussion of exocomet detection in Sect. \ref{s:discussion}. We finally summarize the study in Sect. \ref{s:summary}

\section{Mission parameters}
\label{s:mission}

This feasibility study concerns imaging of exocomets with \textit{LIFE}. The design of \textit{LIFE} is under active evaluation \citep[e.g.,][]{hansen2022}, but here we will use the same reference case design of \textit{LIFE} as in \citet{quanz2022}. The design involves four telescope units, which as a baseline in this study will be assumed to each be 2 m in diameter. The unit telescopes relay their received signals to a beam combiner unit for interferometric purposes. \textit{LIFE} will offer a normal interferometric imaging mode, but it will also offer a nulling interferometry mode, which allows to effectively cancel out the light from a bright star through destructive interference, whilst conserving light signals arising off-axis from the star, such as from planets, or indeed, comets. The nulling interferometry mode offers the best contrast at small separations from bright stars, which means that it is the natural mode to use for exocomet observations. The telescope spacecraft are foreseen to move in a rotation-like manner to optimize UV-plane coverage.

\begin{figure}[htb]
\centering
\includegraphics[width=9cm]{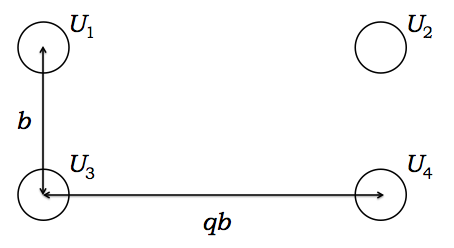}
\caption{Schematic representation of the \textit{LIFE} nulling configuration, where the four unit telescopes $U_{\rm 1}$ through $U_{\rm 4}$ are plotted as circles. The configuration is rectangular with a short axis $b$ and a long axis $qb$. During observations, the configuration rotates in the plane of the targeted sky patch.}
\label{f:config}
\end{figure}

The reference case design includes a low-resolution spectroscopic mode (R$\sim$20) spanning a wavelength range of 4-18.5 $\mu$m. It therefore captures a large fraction of the thermal radiation emitted from a temperate object such as an exocomet visiting the inner parts of a stellar system \citep[e.g.,][]{shanklin2000}, and also incorporates potentially important spectral features such as the silicate feature around 10~$\mu$m. The interferometric baseline will be adjustable in any design, but is foreseen to mainly operate in the range of tens to hundreds of meters. The nulling mode of \textit{LIFE} is foreseen to operate in a rectangular configuration (see Fig. \ref{f:config}), where the shorter axis is called the `nulling baseline' with length $b$, and the longer axis is called the `imaging baseline' with length $qb$ where, following \citet{dannert2022}, $q = 6$. In this study we will simply assume a nulling baseline of $b = 25$~m (unless stated otherwise). This gives an effective size of a spatial resolution element of a few tens of mas within the \textit{LIFE} operating wavelength range. The field of view (FOV) for interferometric observations is set by the beam size of the individual telescope units, $\lambda / D$. Thus, if $D = 2$~m and $\lambda = 10$~$\mu$m, the FOV is 1 arcsec wide. For $S/N$ (signal-to-noise ratio) and integration time calculations, we use the LIFEsim software \citep{dannert2022}, which simulates the nulling interferometry mode of \textit{LIFE} to produce realistic performance estimates.

\section{Observing exocomets}
\label{s:observability}

All observational evidence for exocomets that exist to date originate from indirect measurements. The first evidence arose from the detection of temporally varying narrow spectral features superimposed on the broad spectral lines of the primary star in the $\beta$ Pic system \citep{ferlet1987,lagrange1988}. This was interpreted as being due to exocomets on eccentric orbits, passing so close to the star that they evaporate -- so called Falling Evaporating Bodies \citep[FEBs; see][]{beust1990}. Subsequently, the same type of features have been identified in the 49~Cet \citep{montgomery2012} and HD~172555 \citep{kiefer2014a} systems. A list of a few tens of more tentative FEB detections can be found in e.g. \citet{strom2020}. Further concrete evidence for comets around $\beta$~Pic came with the discovery of a cometary transit in the system \citep{zieba2019}. Similar transit events had previously been identified in the KIC~3542116 system \citep{rappaport2018}. While the exact morphology of a cometary transit depends on orientation \citep{lecavelier1999}, the shape can generally be described as asymmetric, usually with a relatively steep flux drop initially, followed by a smoother rise back toward the baseline. Such a transit is not likely to be mixed up with any alternative known mechanisms, as long as the $S/N$ is reasonably high. Subsequently, analysis of \textit{TESS} \citep{ricker2015} data has revealed several tens of additional cometary transits in the $\beta$ Pic system, which has even made it possible to derive a size distribution for the underlying comet population \citep{lecavelier2022}.

In this paper, we focus on observability of comets through direct imaging, which has so far not been accomplished. In the following, we will discuss the general factors of importance for exocomet imaging, which will be applied particularly to the case of $\beta$ Pic in Sect. \ref{s:frequency}.

A famous feature of comets in the Solar System is the long cometary tail, and for this reason it may seem natural to consider exocomets as having extended flux distributions. However, the tail is typically very faint relative to the head of the comet, and despite the fact that \textit{LIFE} would have an excellent spatial resolution at infrared wavelengths compared to most existing facilities, the resolution is still very coarse relative to the sizes of bodies in a stellar/planetary system. The density of the tail downstream from the comet nucleus can be described as an exponential drop-off $\rho = ce^{-{\lambda}x}$ \citep[adapted from][]{zieba2019}, where $c$ is a normalization constant and $x$ is a spatial coordinate, such that $1/\lambda$ is the scale length of the tail. For the deepest and most well-modeled transiting exocomet, assuming a characteristic star-comet separation of 1 au, \citet{zieba2019} find a scale length of $\sim 2 \times 10^8$ m, or approximately 10\% of the stellar diameter\footnote{There is a typo in the corresponding calculation in \citet{zieba2019} where the unit is written as `km' instead of `m', but a recalculation of their input values and the context of their discussion both confirm that the correct and intended value is $2 \times 10^8$ m.}. This in turn corresponds to $\sim$0.07 mas at the distance of $\beta$ Pic, and even for the closest possible system of $\alpha$ Cen, it is only 1 mas. This is more than an order of magnitude smaller than the spatial resolution for the reference case \textit{LIFE} design, as outlined in Sect. \ref{s:mission}. For all practical purposes in this study, we can therefore consider all exocomets as essentially perfect point sources.

The spectral energy distributions of exocomets cannot be determined from transit observations, so for this aspect we need to rely on information from comets in the Solar System \citep[e.g.,]{gehrz1992}. For example, the flux distribution of Hale-Bopp \citep{shanklin2000} during its active cometary phase could to first order be formulated as a blackbody plus a strong silicate emission feature around 10 $\mu$m \citep{williams1997}. The silicate emission is however not a ubiquitous feature in cometary spectra \citep[e.g.,][]{hanner1994,hanner1996}. Hence, except where otherwise stated, here we will simply model the exocomet spectral energy distribution (SED) as a blackbody. This should be a conservative estimation, since silicate emission would make a comet brighter in the wavelength range of maximum sensitivity for \textit{LIFE}. The SED then depends strictly on the temperature and total effective optical surface of the dust in the coma. Following the modeling in \citet{zieba2019}, we assume that the dust is optically thin. The total optical surface area can then be determined simply from the depth of the transit light curve. We assume that the dust thermalizes quickly enough that the temperature of the grains can be determined from the equilibrium temperature $T_{\rm eq}$ at the instantaneous star-comet distance $r$ at any given time: 

\begin{equation}
T_{\rm eq} = \left( { L_{*}(1-A) \over 16 \pi \sigma_{\rm SB} r^2 } \right)^{1/4}
\end{equation}

where $\sigma_{\rm SB}$ is the Stefan-Bolzmann constant and $L_{*}$ is the luminosity of the star. In order to calculate $T_{\rm eq}$ we need an estimate for the albedo $A$ of the dust. Here we again need to rely on Solar System comets. \citet{kolokolova2004} compile a sample of measured cometary dust albedos, from which we can deduce (for nonextreme phase angles), a range between $\sim$0.1 and $\sim$0.4, with a characteristic mean albedo of 0.25 \citep[e.g.,]{gehrz1992,mason1998,mason2001}. This is similar to the albedo of Earth at $\sim$0.3. Putting this information together, we can note that the median transit depth for the distribution of observable transiting comets in the $\beta$ Pic system is 174 ppm \citep{lecavelier2022}. For reference, given the radius of $\beta$ Pic of 1.5~$R_{\sun}$ \citep{zwintz2019}, an Earth-sized planet would exhibit a transit depth of approximately 40 ppm in the system. Hence, a typical comet in the observable transiting population around $\beta$~Pic has a total optical surface area much larger than that of an Earth-sized planets. Furthermore, at equal separations from the star (characteristically $\sim$1 au for the comet population at time of transit), the comets would also have a similar equilibrium temperature, or perhaps slightly higher (due to the slightly lower albedo), than an Earth-like planet. Since \textit{LIFE} is scoped to be able to image Earth-like planets in and around the habitable zones of stars out to $\sim$20 pc, it naturally follows that it is also well scoped to image large exocomets such as those observed around $\beta$ Pic at a distance of 19.7 pc \citep{gaia2016}.

\section{Potential target systems}
\label{s:targets}

\subsection{$\beta$ Pictoris}
\label{s:betapic}

\begin{figure}[htb]
\centering
\includegraphics[width=9cm]{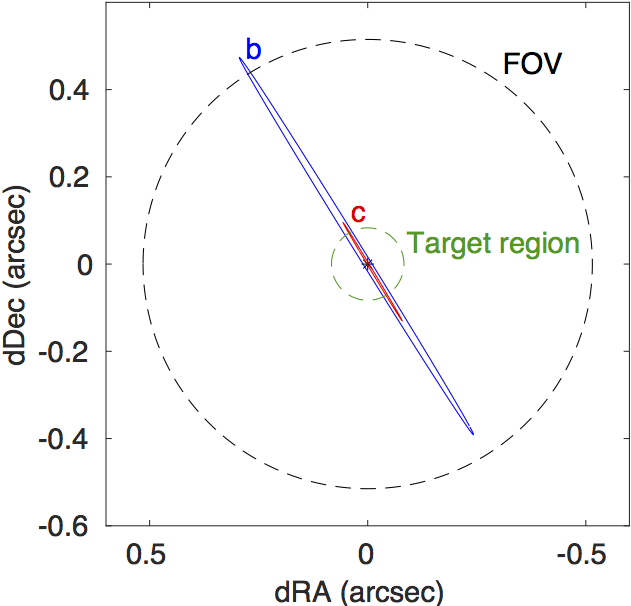}
\caption{Extent of the $\beta$~Pic planetary system relative to the $\sim$1 arcsec FOV of \textit{LIFE} at 10~$\mu$m (black dashed circle), as seen from Earth. The orbits of the planets \citep{nowak2020} are plotted in blue for the outer planet 'b' and red for the inner planet 'c'. The inner green dashed circle marks the region inside of 1.6~au which is of primary interest for exocomet studies, since coma activity may be very limited outside this range. The edge-on disk in the system extends 200 times beyond the \textit{LIFE} FOV.}
\label{f:betamap}
\end{figure}

$\beta$ Pictoris is an A6V-type \cite{gray2006} star with an estimated mass of 1.83 $M_{\rm sun}$ \citep[e.g.,][]{brandt2021} and luminosity of 8.5~$L_{\rm sun}$ \citep{zwintz2019}. $\beta$ Pic is one of the most well-studied young nearby stars, and is the defining member of the $\beta$ Pictoris moving group \citep[BPMG, see][]{zuckerman2001}, which has a collectively determined age of approximately 24$\pm$3 Myr \citep{bell2015}. Due partly to its youth and proximity, $\beta$ Pic hosts a particularly prominent debris disk. It was the first circumstellar disk to ever be spatially resolved \citep{smith1984}, after having been identified through the infrared excess of its host star \citep{aumann1985}. The disk extends out to at least 2000 au \citep{janson2021b} and contains a range of structures and asymmetries \citep[e.g.,][]{kalas1995}, including a prominent warp \citep[e.g.,][]{heap2000,golimowski2006} due to disk-planet interactions. To date, two giant planets have been discovered in the system: $\beta$ Pic b \citep{lagrange2010} at 10.3 au and $\beta$ Pic c \citep{lagrange2019,nowak2020} at 2.7 au; both with masses in the range of $\sim$9~$M_{\rm jup}$ \citep{brandt2021,lacour2021}. The architecture of the system relative to the approximate field of view of \textit{LIFE} is shown in Fig. \ref{f:betamap}.

Given the large amount of planetesimals available in the $\beta$ Pic system as evidence by the bright debris disk, and the presence of at least two massive planets to dynamically stir the disk, there are ample opportunities for cometary activity in the system. Indeed, as we have seen in Sect. \ref{s:observability}, $\beta$ Pic is the system where exocometary activity is most concretely proven, and best characterized. The presence of transiting comet data, in particular in this system is what enables us for the first time to make informed estimations about the prospects for directly imaging exocomets. This is the reason for why we use $\beta$ Pic as the baseline case in Sect. \ref{s:frequency}, which is then the basis for a broader discussion of exocomets in nearby systems in Sect. \ref{s:discussion}.

\subsection{$\epsilon$ Eridani}
\label{s:epseri}

At a distance of only 3.220$\pm$0.001 pc \citep{gaia2016}, $\epsilon$ Eri is one of the nearest stars to us, especially in the context of relatively Sun-like stars. Its spectral type is K2V \citep{keenan1989}, and it has an estimated mass of 0.78~$M_{\rm sun}$ \citep{carrion2021} and an age in the range of $\sim$400--800~Myr \citep{janson2015}. It hosts a prominent debris disk \citep{greaves1998}, which appears to be arranged in several distinct belts \citep{backman2009}. While $\epsilon$ Eri has a high chromospheric activity, which has made precise radial velocity (RV) studies challenging, a planet candidate named $\epsilon$ Eri b has been reported in RV observations of the system \citep{hatzes2000}. The exact orbital parameters have been highly uncertain, with early results implying a highly eccentric orbit, but long-baseline campaigns are converging toward a 3.5 au planet with a mass of $\sim0.7$~$M_{\rm jup}$ and a low eccentricity \citep[e.g.,][]{llop2021}. A (multi)ringed structure such as the one seen in the $\epsilon$ Eri disk can imply the presence of additional planets \citep[e.g.,][]{quillen2006} on wider separations than $\epsilon$ Eri b, although direct imaging studies have not yet detected any such planets, setting upper limits of $\sim$1~$M_{\rm jup}$ from 3 au and outwards \citep[e.g.,][]{janson2015,mawet2019}.

Much like $\beta$ Pic, $\epsilon$ Eri therefore hosts a combination of a bright debris disk, indicating a large supply of planetesimals, and one or several wide giant planets, indicating the possibility for substantial dynamical stirring of the planetesimal population. It thus may be one of the most promising targets for exocomet searches. Unlike $\beta$ Pic, it does not yet feature any direct evidence of such comets, but on the other hand, its extreme proximity offers many advantages for the purpose of detecting exocomets, if such comets do exist there. 

\subsection{Fomalhaut}
\label{s:fomalhaut}

Fomalhaut hosts one of the most prominent debris disks in the Solar neighborhood \citep{kalas2008}. It is an A4V-type \citep{gray2006} star with a mass of 1.92 $M_{\rm sun}$ \citep{mamajek2012} and is located at a distance of 7.7 pc \citep{vanleeuwen2007}. A point source known as Fomalhaut b has been repeatedly seen in several epochs of optical imaging with the \textit{Hubble} telescope \citep[e.g.,][]{kalas2008,kalas2013}. The visible-light point source has no infrared counterpart, which excludes the possibility that it originates from a planetary surface \citep{janson2012}, but instead, it must be scattered light from a local concentration of dust. There could still be a planet at the center of the dust cloud \citep{pearce2021}. However, detailed measurements of the source properties in \textit{Hubble} data has shown that the point source appears to in fact be mildly extended, and expands with time \citep{gaspar2020}. It also appears to be possibly accelerating outwards as a result of radiation pressure. This implies that the dust cloud is transient, and caused by the destruction of a planetesimal-sized body (of order $\sim$100 km) through a catastrophic collision with another planetesimal \citep{lawler2015} or through tidal disruption due to gravitational interaction with one of the suspected planets in the system \citep{janson2020}.

Regardless of the underlying mechanism, the disruption of a planetesimal on an eccentric orbit, and the fact that it was detected serendipitously, indicates a large frequency of cometary-type bodies in the Fomalhaut system \citep{lawler2015}. The favorable proximity of the system, combined with this high expected cometary activity, makes Fomalhaut a compelling target for exocomet searches. 

\subsection{AU Mic}
\label{s:aumic}

AU Mic is an M1V-type star \citep{keenan1989} with a mass of $\sim$0.5~$M_{\rm sun}$ \citep{plavchan2020}, and is possibly the youngest star within 10 pc, given its distance of 9.7 pc \citep{gaia2016}. AU Mic is established as a member of the BPMG \citep[e.g.,][]{malo2014}, just like $\beta$ Pic, which results in an estimated age of 24$\pm$3 Myr \citep{bell2015}. Also much like $\beta$ Pic, AU Mic hosts a very prominent edge-on debris disk \citep{kalas2004}, with interesting substructures. In particular, several dust clumps have been identified in the disk that move outwards from the star \citep{boccaletti2015,boccaletti2018}, possibly implying recurring dynamical phenomena in the inner parts of the system. Additionally, there are two known transiting planets in the system, AU Mic b and c \citep[e.g.,][]{plavchan2020,szabo2022}, at semi-major axes of 0.06 and 0.11 au and masses of 12 and 22 $M_{\rm Earth}$, respectively \citep{zicher2022}, further emphasizing opportunities for dynamical stirring and exocometary activity in the system. AU Mic therefore constitutes another concrete example of the a priori most interesting systems to study in the context of direct imaging of exocomets.

\subsection{Other systems}
\label{s:other}

Beyond the individual examples that are highlighted above, there are a large number of systems in the Solar neighborhood where exocomet searches might be expected to be particularly fruitful. Debris disk targets constitute a promising object class in this context, since the dust observed in such disks is produced continuously through planetesimal disruptions - i.e., a debris disk is evidence for the presence of a large number of planetesimal bodies in the system. \citet{rebollido2020} do not find a correlation with debris disks in general for targets with FEB-type comet signatures, but do find a correlation with near-infrared excess.

Within 20 pc, there are 20 systems with known debris disks listed in the Circumstellar Disks Catalog \footnote{\url{https://www.circumstellardisks.org/}}. Aside from the already highlighted systems, noteworthy examples include GJ 581, which hosts a debris disk along with several RV-detected planets \citep[e.g.,][]{bonfils2005,robertson2014}; $\eta$~Crv, which hosts a particularly bright warm debris disk, implying planetesimal activity around the habitable zone region \citep[e.g.,][]{defrere2015}; and Vega, which hosts a debris ring with a wide gap inside of it \citep{hughes2012}, similar in structure to systems like Fomalhaut or $\epsilon$ Eri.

Another path toward candidate exocomet imaging systems is if there are other indicators for exocomets, such as spectral lines that can be associated with FEBs, or exocomet transits. As discussed in Sect. \ref{s:observability}, some such targets are already known, although they are not always suitable for observations with \textit{LIFE}. For example, the exocomet transit target KIC~3542116 \citep{rappaport2018} is much too distant at $\sim$260 pc to be a feasible target for imaging around the $\sim$1 au range, and the FEB target HD~172555 \citep{kiefer2014a} is in principle feasible, but not ideal at 28.8 pc \citep{gaia2016}. Both \textit{TESS} \citep{ricker2015} and \textit{CHEOPS} \citep{benz2021}, and in the future \textit{PLATO} \citep{rauer2014} are sensitive to exocomet transits around nearby stars, and may thus provide additional proven exocometary systems well ahead of the \textit{LIFE} mission. 

Beyond targeted exocomet searches, exocomets could conceivably be detected serendipitously as part of the larger \textit{LIFE} program. \textit{LIFE} will likely spend a large fraction of its mission time studying relatively Sun-like main-sequence stars, with a sufficient contrast to detect both terrestrial planets at various stages of their evolution and exocomets. It is not possible to make an accurate estimation of how common observable exocomets will be in such systems, not least because the Solar System may not necessarily be representative of a typical main-sequence star in this regard. However, as a rough estimation, we can consider the fact that in the Solar System, so-called `great comets' with comparable sizes to what has been inferred from the exocomet distribution in \citet{lecavelier2022} occur approximately once per decade. Since the \textit{LIFE} mission will extend for several years, and perhaps up towards a decade, it could be possible for individual exocomet events to be discovered in the main sample through sheer serendipity, but most likely not at any high rate.

\section{Simulations}
\label{s:casestudy}

Here we present the outcomes of a series of \textit{LIFE} observation simulations for the specific case of $\beta$ Pic, as a baseline case for considering exocomet observability.

\subsection{Frequency and detectability of comets}
\label{s:frequency}

As mentioned previously, 30 exocomet transits have already been identified in the $\beta$~Pic system. This allows us to both assess their individual prospects for being directly imaged, as outlined in Sect. \ref{s:observability}, but also to estimate the rate at which they might be expected to occur. 156 days of \textit{TESS} observations revealed 30 individual cometary transit events, with absorption depths between 107 ppm and 1963 ppm. For interpreting and simulating the corresponding size of the effective optical surface, it is useful to formulate this surface in terms of the radius (in units of Earth radii, $R_{\rm E}$) that a circular disk with the same effective surface area as the cometary dust coma would have. For the observed $\beta$~Pic exocomet population, this gives effective radii between 1.6~$R_{\rm E}$ and 7.0~$R_{\rm E}$.

\begin{figure}[htb]
\centering
\includegraphics[width=9cm]{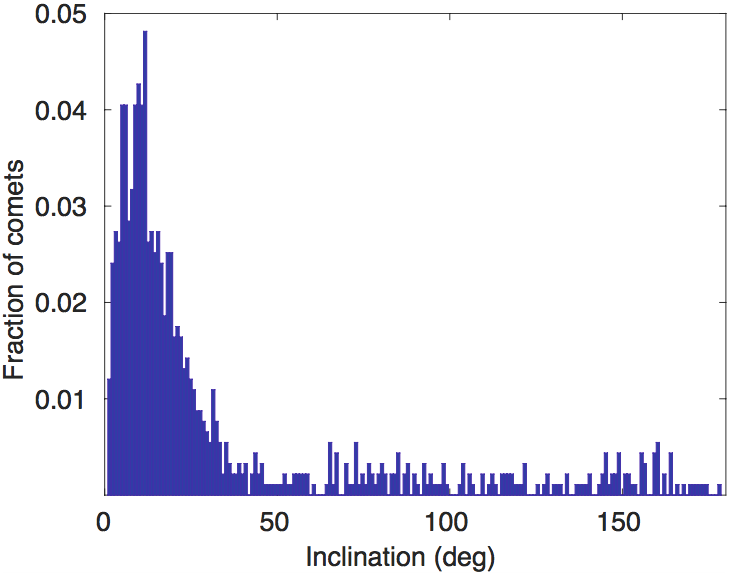}
\caption{Histogram of the inclinations of $<$100~au comets in the Solar system. While low inclinations are generally statistically favored, the spread of the distribution is quite wide. Thus, if the comets in a system like $\beta$ Pic share a similar orbital distribution, then only a small fraction of comets will transit as seen from any given orientation.}
\label{f:inchist}
\end{figure}

While exocometary transits can only be observed at a narrow range of orbital inclinations, direct imaging can reach exocomets at any inclination. Thus, to estimate the frequency of exocomets around $\beta$~Pic amenable to direct imaging, we need to make some assumption about the inclination distribution of the comets. Since the $\beta$~Pic circumstellar disk and planets are all encompassed by a narrow inclination distribution which is close to edge-on \citep[$\sim$85--88 deg in different parts of the warped disk, e.g.][]{kalas1995,ahmic2009,janson2021b}, it might be tempting to assume that the exocomets follow a similar distribution. In this case, a relatively large fraction of the exocometary population may already be captured in the transit monitoring. However, it is important to note that assuming such a distribution would give drastically incorrect results for comets in the Solar system. We have compiled a list of all object classified as comets with semi-major axes $<$100 au in the Small-body database\footnote{\url{https://ssd.jpl.nasa.gov/tools/sbdb_query.html}}, and show the distribution of inclinations in Fig. \ref{f:inchist}. The observed transiting comets are expected to typically reside at $\sim$1~au from the star when the transit occurs \citep[e.g.,][]{zieba2019}. At this separation, a transit occurs if the comet has an inclination within $\pm$0.48 deg from perfectly edge-on. Hence, the observable inclination band is 0.96 deg. We therefore use 0.96 deg as the sampling size for the distribution in Fig. \ref{f:inchist}, and normalize by the total number of objects, in such a way that the y-axis reflects the fraction $\phi$ of Solar systems comets that would be detectable in transit if viewed from a particular direction (as represented by the x-axis). Although there are many more comets close to the ecliptic plane than far from it, the peak of the distribution occurs quite far ($\sim$10--20 deg) from the ecliptic, and the distribution is so broad than no particular $\sim$0.96 deg inclination interval captures any substantial fraction of the comets. The few most favorable directions capture up to $\sim$4-5\% of the comets, and the least favorable directions much less than 1\%. If we assume that $\beta$~Pic has a similar cometary inclination distribution as in the Solar system, it seems reasonable to assume $\phi = 0.01$ as a characteristic value for the fraction of comets that exhibit transits. We therefore adopt this value as a baseline, but at the end of the section we will also note the implications of adopting other values of $\phi$.

In order to evaluate the detectability of any individual exocomet, we use the LIFEsim \citep{dannert2022} software, which simulates the achievable signal-to-noise ratio $S/N$ for a given integration time $t_{\rm int}$, and a range of stellar, observatory, and ``planetary'' (in our case, cometary) parameters. The required stellar system parameters are the effective temperature $T_{*}$, radius $R_{*}$, distance $d_{*}$, ecliptic latitude $l_{*}$, and exozodiacal level $z_{*}$ in units of Solar system zodiacal brightness (zodi). We use $T_{*} = 8100$ K and $R_{*} = 1.5 R_{\rm sun}$ from \citet{zwintz2019}, and $d_{*} = 19.7$ pc, and $l_{*} = 4.98$ rad from the NASA exoplanet archive based on \textit{Gaia} measurements \citep{gaia2016}. The exozodiacal level in the $\beta$ Pic system is unknown, so we run simulations with $z_{*} = 1$, $z_{*} = 10$, $z_{*} = 100$ and $z_{*} = 1000$ zodi. Since $\beta$ Pic has such a bright debris disk in general, it is likely that the zodiacal disk is also quite bright, so we use $z_{*} = 100$ zodi as a baseline value.

The observatory parameters refer to the spectral resolution, the sizes of the unit telescopes, and the interferometric baseline between the telescopes. For spectral resolution, we selected the default setting, which is $R = 20$. For telescope aperture sizes, a frequently used reference value for the mission is 2 meters, which we choose as a characteristic size also in this study. For interferometric baseline, we experimented with different lengths, and settled for 25 meters as a main option, since this offers good sensitivity across the most interesting range for detecting exocomets (projected separations of $\sim$0.5--1 au).

The relevant planetary/cometary parameters, assuming a blackbody flux distribution, are the projected separation, the effective temperature, and the effective radius. For studying a range of possible parameter values, we implemented a 2-dimensional grid system, where effective radii ranged from 2~$R_{\rm E}$ to 7~$R_{\rm E}$ in steps of 1~$R_{\rm E}$, and projected separations ranged from 20 to 80 mas in steps of 20 mas. To calculate an effective temperature for each grid point, we assumed a template comet with nearly identical orbital parameters as Halley's comet (as listed in the Small-body database), except that the orbital period was scaled according to the $\beta$ Pic system mass based on a fixed semi-major axis of 17.834 au, and the inclination was set to 60 deg in order to reflect a median inclination for a highly scattered (effectively uniform) hypothetical exocometary distribution. We simulated the orbit based on these parameters, and with a sampling frequency of 1 day, we calculated the effective temperature (based on the instantaneous star-comet separation, see section \ref{s:observability}) and projected separation at each sampled time-step. Choosing (arbitrarily) the post-periastron section of the orbit, we then have an estimate for the effective temperature as function of projected separation for comets on Halley-like orbits, which we interpolate to sample at the desired grid points of 20 to 80 mas, with a step size of 20 mas.

\begin{figure*}[htb]
\centering
\includegraphics[width=15cm]{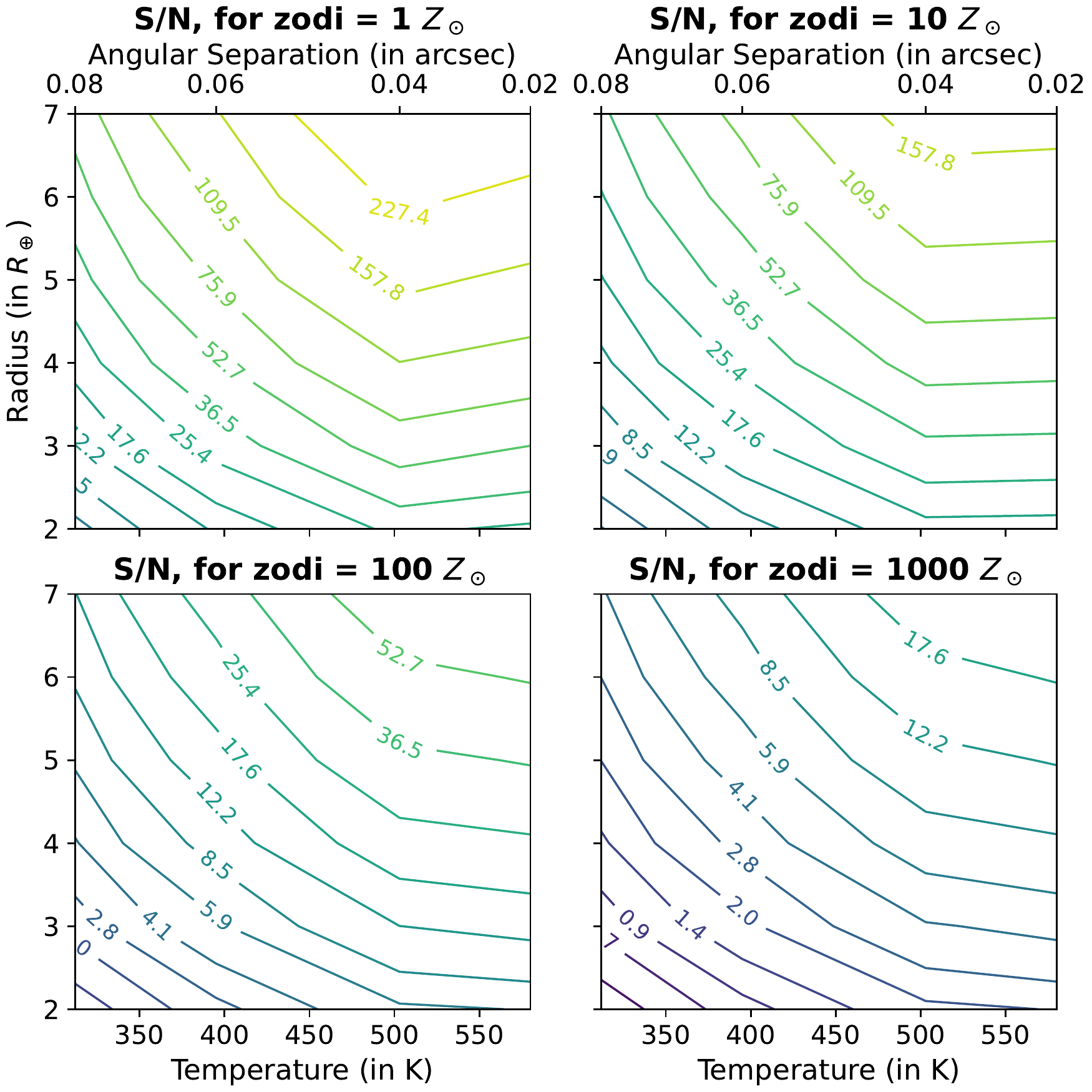}
\caption{$S/N$ values (color scale) as function of temperature and effective coma radius for simulated exocomets in the $\beta$ Pic system, in 10 h of observing time. Upper left: A case with low zodiacal contamination (1 zodi). Upper right: A higher level of 10 zodi. Lower left: The baseline value of 100 zodi. Lower right: A very high level of 1000 zodi.}
\label{f:snrcontours}
\end{figure*}

\begin{figure*}[htb]
\centering
\includegraphics[width=18cm]{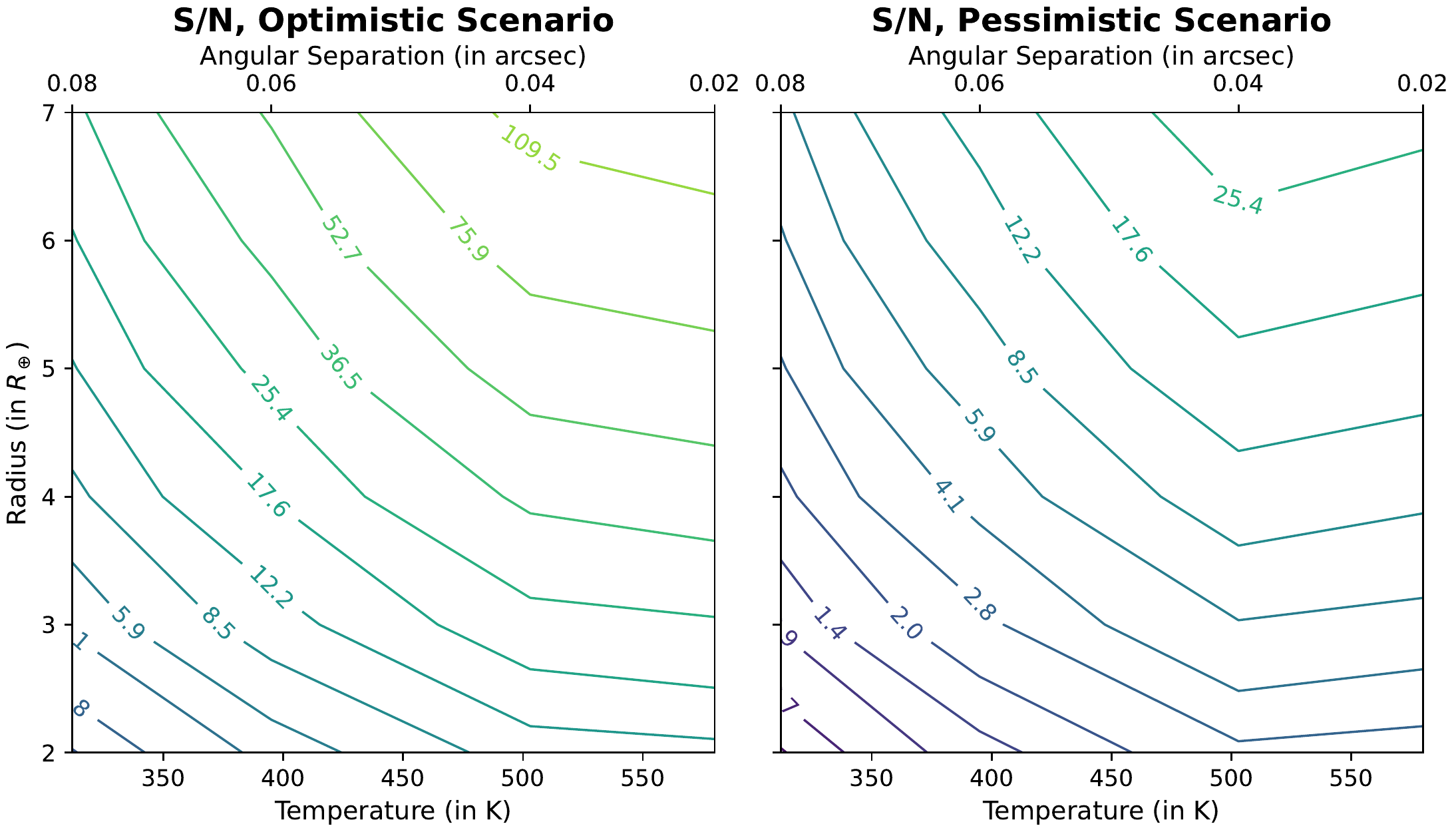}
\caption{Same as Fig. \ref{f:snrcontours}, but varying the sizes of the unit telescopes from the baseline value (2~m diameter). The observing time in each case is 10 h. Left: 3.5~m apertures. Right: 1.0~m apertures.}
\label{f:morecontours}
\end{figure*}

The $S/N$ values calculated with LIFEsim across the separation-radius grid with our default parameters are shown in Fig. \ref{f:snrcontours}. It is immediately clear that a wide range of cometary bodies would be detectable at high significance in the $\beta$~Pic system in 10 hours of integration time. Our baseline setting has $z_{*} = 100$ zodi, but we also plot calculations for other exozodiacal levels in the same figure. As expected, lower exozodiacal levels give even higher $S/N$ values in the same amount of time, but even with substantially higher levels ($z_{*} = 1000$ zodi), detection remains possible. In Fig. \ref{f:morecontours}, we show alternative cases with default stellar and cometary properties, but varying the telescope unit aperture size, to account for the boundary scenarios of possible mission sizes in \citet{quanz2022}. The $S/N$ scales approximately proportionally to unit telescope size, as expected. We also checked the detectability impact of the interferometric baseline length. The overall impact changing the interferometric baseline is minor, but the sensitivity increases at smaller separations and decreases at larger separations for longer interferometric baselines, and vice versa.

With $S/N$ values across the full size-separation grid in hand, and prior knowledge of the frequency and effective size distribution of the $\beta$ Pic comet population from previous works, we can now evaluate how commonly observable exocomets can be expected to occur. We do this by once again assuming the same Halley-like template orbit as before, considering the half orbital arc that occurs after periastron for simplicity. We then generate 1000 simulated exocomets where each comet is randomly assigned a size from the sample of known exocomet effective sizes generated from the absorption depth distribution in \citet{lecavelier2022}. Out of the 30 exocomets listed in \citet{lecavelier2022}, We only use the 18 exocomets with effective sizes $>$2~$R_{\rm E}$ for representing the size distribution, since smaller comets are never observable with $S/N > 5$ in 10 hours of observations for $\beta$~Pic under the baseline assumptions. For each comet, we interpolate the $S/N$ grid at the locations corresponding to the comet's size and the projected separations given by the simulated orbit (which is sampled with a 1-day cadence). We calculate for how many days each exocomet is observable, where $S/N > 5$ is the threshold for what is considered observable. The average duration of observability is 83.5 days across the simulated orbital half-arc, and therefore approximately two times that number for a full arc. The average duration of observability (in days) multiplied by the number of relevant exocomets that enter the observability zone per day gives the number of exocomets that are observable at any given time. The number of entering comets can be calculated by considering the fact that the 18 selected transiting cometary events occurred over a \textit{TESS} observing window of 156 days in total. Hence, 0.115 sufficiently large transiting comets enter the window each day, so if the fraction of comets that transit is $\phi = 0.01$ as discussed above, 11.5 potentially imageable comets enter the same window on a daily basis. In total, there is then an average of $\sim$960 exocomets that are bright enough to be observable with \textit{LIFE} at any random point in time. 

This remarkably high number would in fact mean that the observations would be heavily confusion limited, both since the number of exocomets would far exceed the number of resolution elements inside of $\sim$80 mas, and since \textit{LIFE} has a limited effective UV-plane coverage, making it difficult to accurately reconstruct a complex mix of point sources, compared to reconstructing a field containing just one or a few point sources. We will discuss this in more detail in Sect. \ref{s:crowding} and outline the implications in a broader context in Sect. \ref{s:discussion}, but as a basic first conclusion, it appears clear that individually detectable exocomets are plentiful in the $\beta$~Pic system, which can probably be seen as constituting an upper envelope for the number density of exocomets around nearby stars. 

The above calculations depend heavily on the value for the parameter $\phi$, which as we have discussed is essentially unknown for $\beta$~Pic, but is around $\sim$1\% if the solar system distribution of comets is representative. Since $\beta$~Pic is younger and its disk and planets more massive, the orbital distribution of comets could also be quite different. For example, the close-in FEB population has been suggested to arise from resonances with the planets in the system \citep[e.g.,][]{beust2000}. This might force subpopulations of the exocomets into relatively narrow orbital families. In the FEB population, two subfamilies have been identified, which have scatters in the argument of periapsis of $\pm$8~deg and $\pm$25~deg respectively \citep{kiefer2014b}. The corresponding scatter in inclination distributions has not been measured from FEB data, and the FEB population is by no means necessarily representative of the exocomet population in total or the \textit{TESS} transiting subpopulation. Hence, accurate clustering properties in the potentially imageable population cannot be predicted, but as a general principle, if the $\beta$ Pic comets happen to be both narrowly distributed and fortuitously aligned toward the line of sight, $\phi$ could be higher and consequently, the number of imageable exocomets could be lower than calculated above. Still, since there are 960 sufficiently bright comets in the observing window if $\phi = 0.01$, it follows that there are 9.6 comets in the same window even in the extreme case where $\phi = 1$, such that every single comet in the system transits. Since all orbits would be edge-on in this circumstance, projection effects would bring those 9.6 comets closer to the effective inner working angle than what would be the case for the template orbit, but the majority of them would still be detectable. Hence, regardless of the specific inclination distribution of comets in the $\beta$~Pic system, the simulations indicate that at least several exocomets would be formally imageable with \textit{LIFE} at any given time.

\subsection{Cometary crowding}
\label{s:crowding}

\begin{figure}[htb]
\centering
\includegraphics[width=9cm]{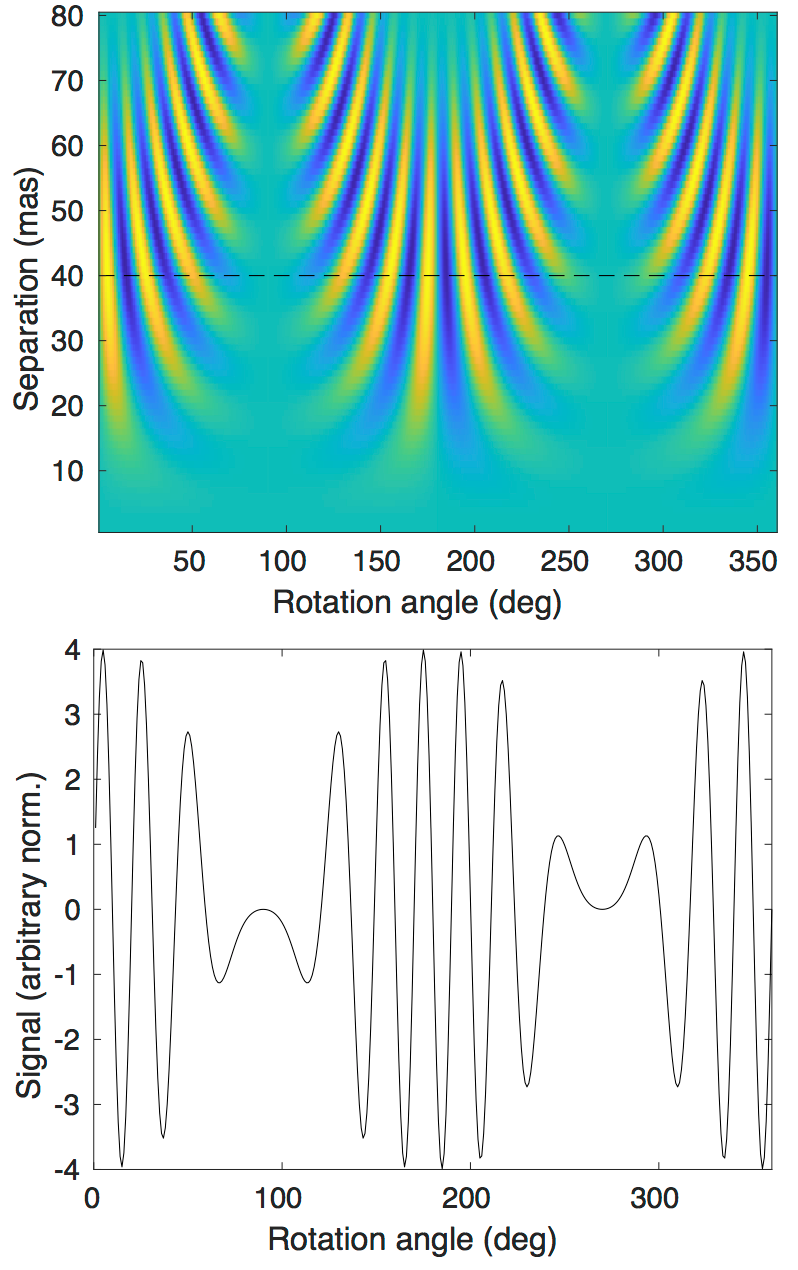}
\caption{Transmission for the nulling interferometric mode used in this study. Upper: Transmission pattern in polar coordinates at 10~$\mu$m of the nulling interferometric configuration used in this study. A comet at a separation of 40 mas would project along the black dashed line as the telescope array rotates. Lower: A cross section of the transmission pattern for a separation of 40 mas (i.e., along the dashed line in the upper panel). A comet residing at a position angle of 0 deg would map the transmission pattern of the black solid line. A comet at a different PA would result in a cyclically phase-shifted pattern of otherwise the same shape. See \citet{dannert2022} for further details.}
\label{f:transmission}
\end{figure}

The configuration and signal combination planned to be used for \textit{LIFE} is described in detail in \citet{dannert2022}. In short, the signal combination from the four unit telescopes nulls out the on-axis light from the parent star, and for the rest of the field, it effectively creates a transmission pattern on the sky with alternating positive and negative regions. As the telescope configuration rotates, so does the transmission pattern, so a planetary or cometary off-axis companion moves in and out of positive and negative areas as seen by the telescope array. An off-axis companion at a given separation and position angle will therefore give rise to a unique modulation of the interferometric signal, as illustrated in Fig. \ref{f:transmission}. Thus, the astrometric and spectrophotometric properties of a single isolated off-axis source can be easily backed out from the nulling signal, even with the very limited $uv$ coverage provided by a four-element nulling array. 

If several off-axis sources exist in the same field, their contributions blend into a mutual interferometric signal, which makes the signal interpretation much more challenging. How to most efficiently back out information about the individual sources in this context is work in progress, and beyond the scope of this study. However, since we can expect some exocometary fields to be crowded, at least in the case of $\beta$ Pic as outlined above, we will discuss an example case of a moderately crowded field here in order to illustrate the impact of such crowding on the interferometric signal.

\begin{figure*}[htb]
\centering
\includegraphics[width=16cm]{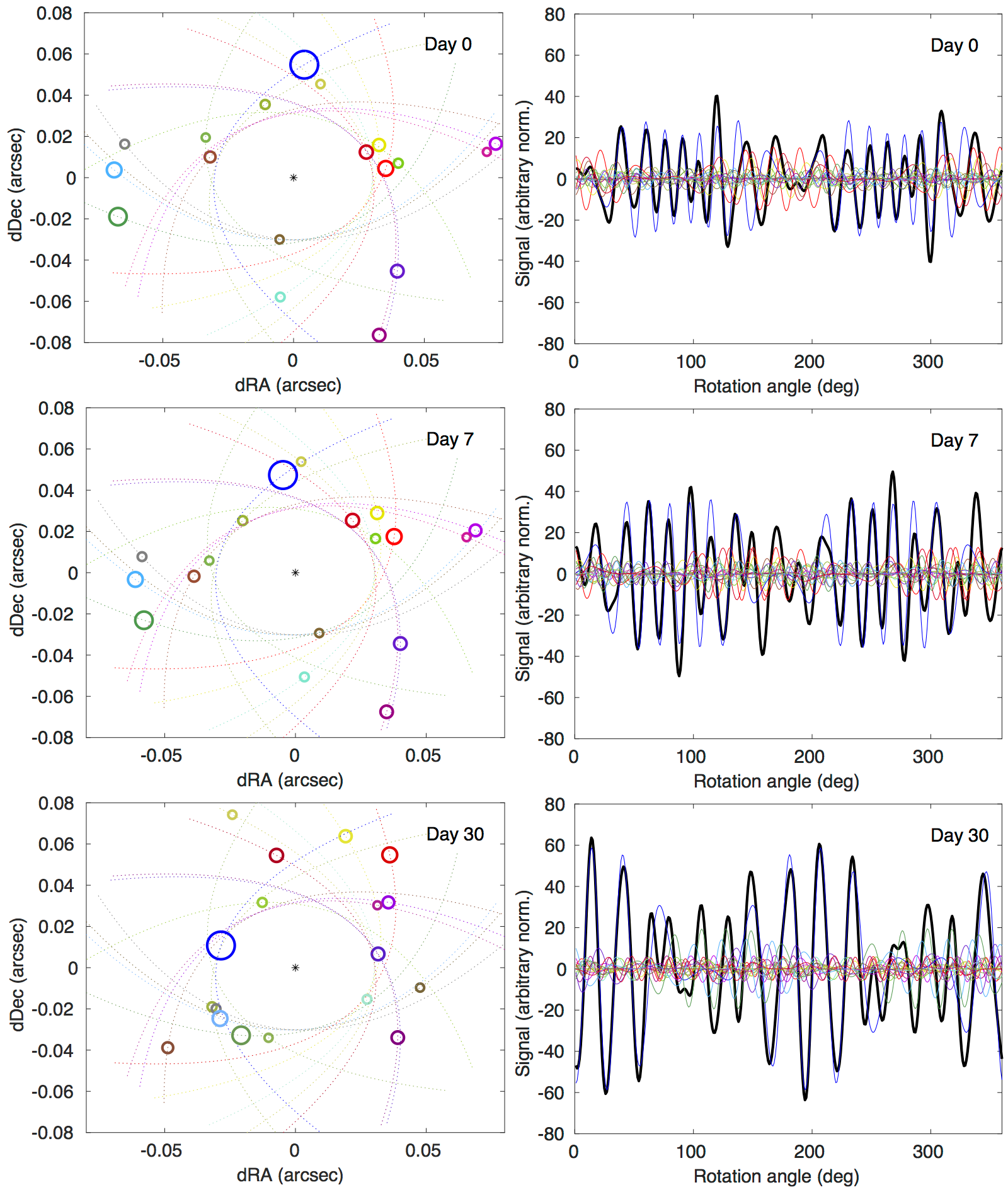}
\caption{Simulations of a crowded exocometary field. The three rows correspond to three different dates. Upper: Day 0. Middle: Day 7. Lower: Day 30. The left column contains maps of the spatial distribution of the exocomets. Each circle is an exocomet, where the radius of the circle is directly proportional to the effective radius of the cometary coma. Each comet has its own color, and the dashed lines of each color corresponds to the orbital track of that particular comet. The right column shows the interferometric signal as a black thick line. For illustrative purposes, the contribution of each individual comet is plotted as a thin line that is color-matched with the corresponding symbol in the left column. The individual colored components cannot be measured directly; they can only be deduced from the total signal and its evolution in time, wavelength, and (if relevant) along different probed interferometric baselines. See the text for discussion.}
\label{f:mapsignal}
\end{figure*}

A simulation with 18 exocomets  at a wavelength of 10~$\mu$m is shown in Fig. \ref{f:mapsignal}. The comets have the same size distribution as the 18 largest comets observed in the $\beta$~Pic system (see Sect. \ref{s:frequency}), and the mass and distance of the star is also set to the same as $\beta$ Pic. In order to make the illustration as clear as possible, we consider a case where all comets share the same orbital plane, and this plane is viewed face-on by \textit{LIFE}. This means that the projected separation is always the same as the physical separation, such that comets that lie closest to the star in image space are also physically the closest to the star, and therefore hotter than the more distant comets. We also assume that all comets have exactly the same orbital parameters, except for the argument of periapsis $\omega$ which is randomly distributed between 0 and 360 deg. The orbits are Halley-like with semi-major axes of 17.834 au and eccentricities of 0.967.

Three different dates are simulated: An original epoch of observation called Day 0, a follow-up observation one week later called Day 7, and another observation 1 month from the original epoch called Day 30. In all three epochs, the total signal is dominated by the largest comet in the field (plotted in blue). This is especially true at Day 30 when the largest comet is also one of the closest to the parent star, and therefore one of the hottest in the field. This increases the total interferometric signal, and also makes it cleaner in the sense that it is dominated by a single source whose properties could probably be quite accurately backed out. On Day 0 and Day 7, the signal would be considerably more degenerate and prone to larger uncertainties. Crowding can therefore impose difficulties in the interferometric signal interpretation, and in the case of the $\beta$ Pic system this is of course emphasized by the fact that, as we have seen in Sect. \ref{s:frequency}, the crowding there could be up to 50 times larger than in this simulated example.  This would mean that even many individual resolution elements are crowded.

In a full signal extraction, it is possible to mitigate the ambiguities resulting from a sparse sample of the $uv$ plane by accounting for all the spectroscopically observed wavelengths simultaneously, instead of just the 10~$\mu$m signal as in this illustrative example. It would also be possible to extend the coverage further by making additional observations with different baseline lengths. Nonetheless, we expect that the extreme crowding expected in the particular case of the $\beta$~Pic system would make it a challenging target for \textit{LIFE}. In the meantime, in the following two sections, we will disregard any possible issues associated with overcrowding of comets, and investigate what information can be acquired from an isolated exocomet, if observed over a handful of epochs.

\subsection{Astrometric constraints}
\label{s:orbit}

Given that exocomets appear to be potentially detectable under certain circumstances, it is relevant to examine to which extent their properties can be constrained, and to which extent they can be distinguished from other possible point sources in the system (such as habitable planets, for example). Here, we focus on constraints related to astrometric measurements of exocomets. 

In \citet{dannert2022}, it is found that the $S/N$-limited astrometric precision of \textit{LIFE} for a point source with $S/N = 9.7$ is 1.5~mas. Beyond a limited $S/N$, astrometric precision is typically subject to calibration uncertainties related to the physical scaling and orientation of the astrometric grid. While such calibration details have not been examined in detail for \textit{LIFE} yet, experience with astrometry from other facilities shows that providing calibration uncertainties at the $\sim$1~mas level or far below is a demonstrably manageable task \citep[e.g.,][]{maire2016,gravity2021}. We therefore assume that the measurements would be $S/N$-limited. Under this circumstance, the astrometric uncertainty $\delta$ typically scales as $\delta \propto (S/N)^{-1}$ \citep[e.g.,][]{lindegren1978}.

In order to assess astrometric performance in a reasonably realistic context, we imagine a scenario in which a point source, suspected as a possible exocomet, is detected in observations acquired at date $T_1$. Follow-up observations are then acquired at two additional epochs ($T_2$ and $T_3$), taken 20 and 40 days later. Each observation contains 10 h of integration time. Astrometric data points are simulated by setting the ``true'' orbital parameters of the comet as corresponding to a Halley-type orbit with a semi-major axis $a_{\rm T}$ of 17.834~au, an eccentricity $e_{\rm T}$ of 0.967, and an inclination $i_{\rm T}$ of 60~deg just like in Section \ref{s:frequency}. The period for a system mass of 1.75~$M_{\rm sun}$ comes out to $P_{\rm T} = 56.9$ years. The time of periastron $T_{\rm T}$ was set to an MJD of 58240.7 days, which in turn was set to three days before the first observation, $T_1 = T_{\rm T} + 3$. The argument of periastron $\omega_{\rm T}$ was set to 111.33~deg, and the ascending node $\Omega_{\rm T}$ to 238.42~deg. The effective size of the exocomet coma was set to 5~$R_{\rm Earth}$.

For each simulated observation date, we extracted the astrometry relative to the central star as $\Delta x_1 = 7.5$ mas (eastward) and $\Delta y_1 = 19.6$ mas (northward) at $T_1$, $\Delta x_2 = 38.2$ mas and $\Delta y_2 = 22.9$ mas at $T_2$, and $\Delta x_3 = 57.6$ mas and $\Delta y_3 = 17.9$ mas at $T_3$. Based on the physical and projected separations at each date, we then calculated $S/N$ values using LIFEsim, acquiring values of 37.2, 29.6 and 15.0, leading to estimated astrometric uncertainties of 0.4 mas, 0.5 mas and 1.0 mas, respectively. Based on these uncertainties, we generated random errors with the corresponding $\sigma$, assuming Gaussian distributed errors, and added the errors of such a random set to the true astrometric positions. This resulted in simulated measurements  $\Delta \tilde{x}_1 = 7.8 \pm 0.4$ mas and $\Delta \tilde{y}_1 = 19.4 \pm 0.4$ mas, $\Delta \tilde{x}_2 = 37.9 \pm 0.5$ mas and $\Delta \tilde{y}_2 = 22.1 \pm 0.5$ mas, and $\Delta \tilde{x}_3 = 57.2 \pm 1.0$ mas and $\Delta \tilde{y}_3 = 18.3 \pm 1.0$ mas.

It is clear from basic considerations that it would be impossible to determine detailed orbital properties of the comet based on the astrometric information given in this simulated example: With only three astrometric epochs, accurate orbital fitting is challenging under any circumstance, but even more so in this context, where the observational baseline only covers a tiny fraction ($\sim$0.2\%) of the full cometary period. Nonetheless, the observations can be expected to provide some useful constraints on the orbit. In particular, the speed and acceleration of a highly eccentric object near periastron are significantly different from (e.g.) a smaller orbit of low eccentricity at the same instantaneous separation. The instantaneous orbital velocity $v_{\rm orb}$ is given by:

\begin{equation}
    v_{\rm orb} = \sqrt{GM_* \left( \frac{2}{r} - \frac{1}{a} \right)}
\end{equation}

where $G$ is the gravitational constant, $M_*$ is the stellar mass, $r$ the instantaneous separation and $a$ the semi-major axis. Thus, for example, an object in a circular 1 au orbit around $\beta$~Pic has an orbital velocity of 40 km/s, while a high-eccentricity object with a semi-major axis of 100 au has an orbital velocity of 57 km/s when it crosses 1 au. Meanwhile, for large $a$ (small $1/a$), $2/r$ becomes the dominant term and $v_{\rm orb}$ becomes very weakly dependent on $a$ -- as a result, the difference in velocity between a 10 au modestly eccentric orbit and a 100 au highly eccentric orbit when crossing 1 au is only $\sim$2\%. As a consequence, while we can expect to be able to distinguish between low and modest-to-high eccentricities in general, we cannot expect to distinguish between modestly high and very high eccentricities, with the limited astrometric information provided in the simulated example.

\begin{figure*}[htb]
\centering
\includegraphics[width=12cm]{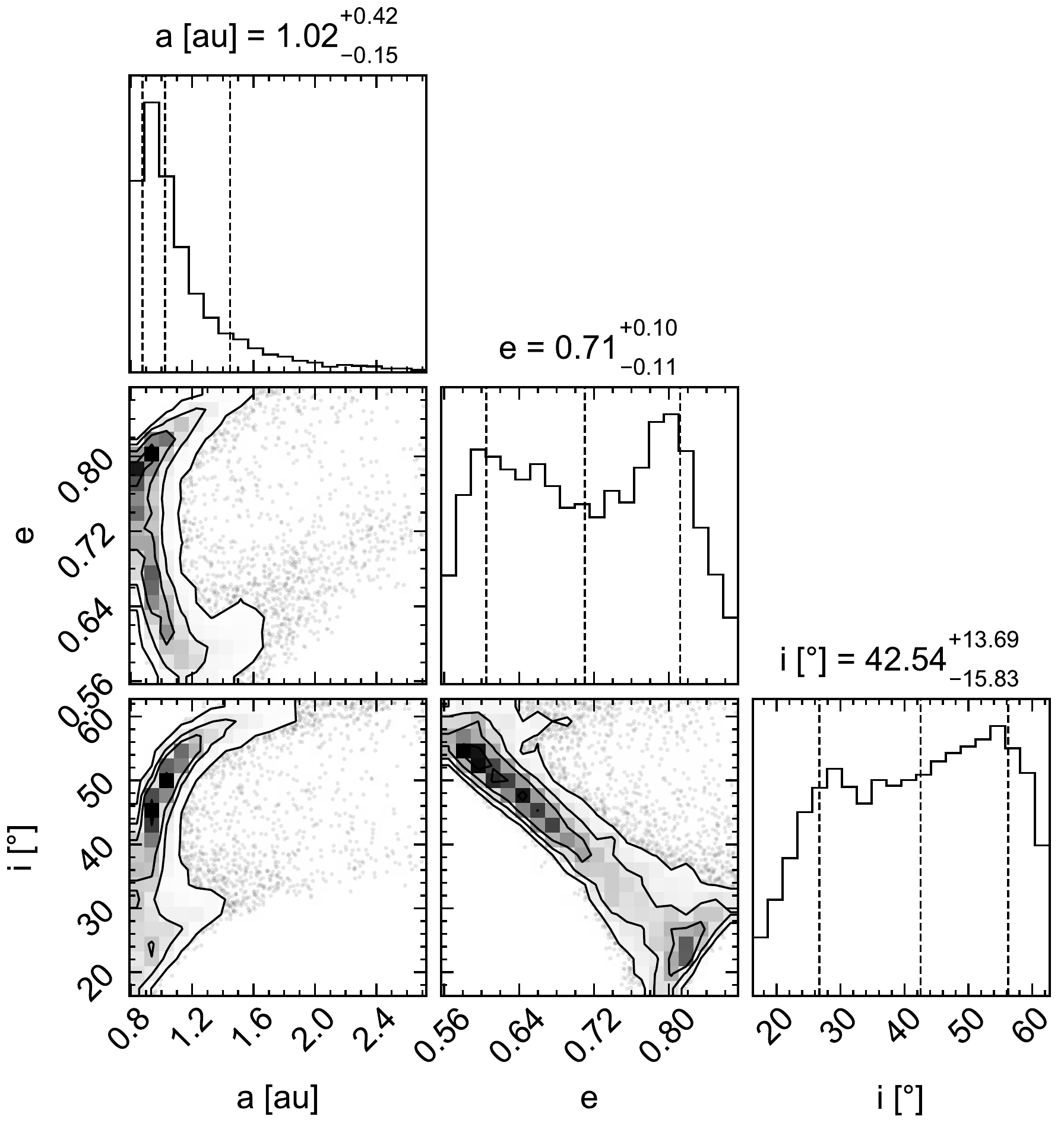}
\caption{Corner plot for the most important orbital parameters from the orbit fitting test described in the text, with a 95\% cut on the axes for the $10^4$ orbital solutions. The true orbital parameters are not well recovered, due in large part to the short baseline probed with the simulated observations.}
\label{f:cornerplot}
\end{figure*}

\begin{figure*}[htb]
\centering
\includegraphics[width=16cm]{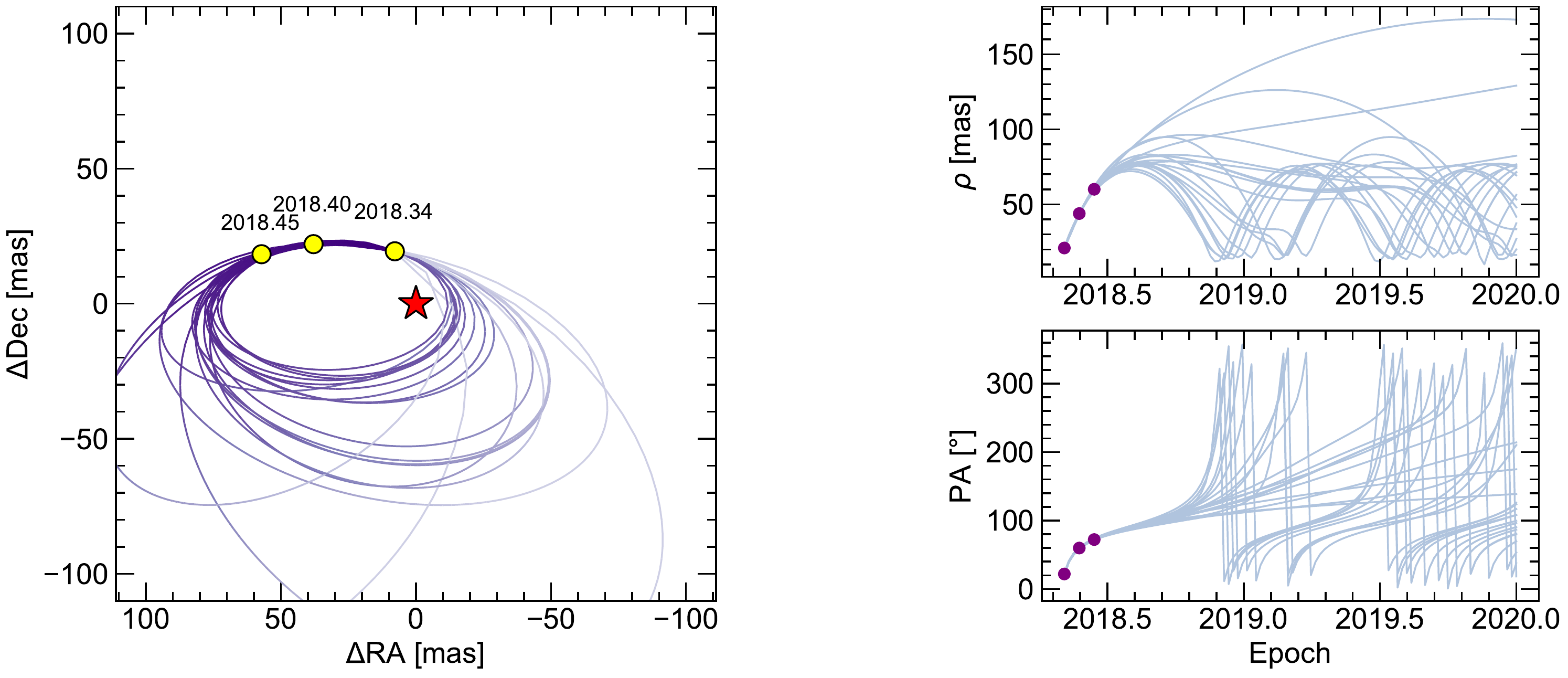}
\caption{20 example fitted orbits (colored lines), randomly sampled from the full population of fitted orbits. The orbits are generally significantly smaller and less eccentric than the input orbit, underlining the fact that the probed baseline is too short to determine exact parameters, although meaningful limits can be obtained (see text for details). Left: The orbital fits in two dimensions. The red star symbol marks the location of the star, and the yellow points mark the locations of the comet at three measured epochs. Upper right: Separation as function of time. Upper left: Position angle as function of time. Purple dots mark the location of the comet at the respective epochs.}
\label{f:orbits}
\end{figure*}

To quantify this, we ran orbital fitting with the OFTI code within the orbitize package\footnote{\url{https://orbitize.readthedocs.io/}} \citep{blunt2020} on the simulated data points. We use a standard set of uninformed priors for the fitting, and run the simulations for $10^4$ orbits. The results of the fitting are shown in Fig. \ref{f:cornerplot}, with example orbits in Fig. \ref{f:orbits}. As expected, the true orbital parameters are poorly recovered. Of particular importance is the disagreement between the true eccentricity of 0.967 and the recovered eccentricity of 0.71$^{+0.10}_{-0.11}$. 

While it is straightforwardly expected that the best-fit eccentricity fails to provide a close match to the true value as argued above, it is perhaps more intuitively surprising that the true value is not recovered within the estimated error bars of the posterior (1.7$\sigma$ deviation). We believe that the primary reason for this is the combination of a very high eccentricity with a phase very close to periastron. This is a priori a highly unlikely configuration, which means it cannot acquire sufficient probabilistic weight to contribute substantially to the posterior. In fact, the only way to observe a Halley-like exocomet is when it is extremely close to periastron, so the reason that we expect Halley-like exocomets to be observable in the first place is closely connected to this strong selection effect, which the fitting algorithm has no good way to account for. In a theoretical setting, we can of course adjust the priors in the fitting procedure to account for this selection effect -- however, this solution is not applicable to the real world, where we have no prior information on the orbital distribution of comets. For example, modestly eccentric bright comets may be much more common in the young and dynamically active $\beta$~Pic system than in the Solar system, and skewing the priors toward extreme eccentricities may yield erroneous results for such a population. 

As a result, we find that it would be impossible to set any stringent upper limit on the eccentricity based on the astrometric data in our simulated example, due to unavoidable biases in the procedure. However, there is no corresponding bias for the lower limit on the eccentricity -- that part of the parameter space is disfavored in the posterior simply on the basis that such orbits are not consistent with the existing simulated data. We thus conclude that the lower limit on the eccentricity is reliable, and that it is therefore possible to at least distinguish orbits of low eccentricity from orbits of intermediate or higher eccentricity, to a good degree of accuracy. Fitted parameters other than the eccentricity generally cannot be relied on in this context -- since the procedure misses the ``true'' solution in terms of the eccentricity, it is likely to falsely estimate the global optima for other orbital parameters as well.

\subsection{Spectroscopic constraints}
\label{s:spectrum}

As mentioned previously, some Solar system comet spectra are featureless blackbody spectra, while others exhibit strong silicate emission. So far in this paper, we have only considered the pure blackbody case. Even in this case, useful information can be extracted from the spectra -- the 4--18.5 $\mu$m spectral range of \textit{LIFE} includes the blackbody peak for temperate exocomets, and the peak wavelength scales inversely with the temperature of the comet. Hence, the temperature of the dust can be derived at each measured epoch. By monitoring the temperature as function of separation over several epochs, it may in turn be possible to derive an estimate for the albedo of the dust in many cases.

However, a perhaps even more compelling set of information could be extracted if the comet shows silicate emission, similar to e.g. comet Hale-Bopp \citep{williams1997} in the Solar system. This would allow one to place constraints on the mineralogy of the dust, Mineralogical studies have been made of dust in a wide range of disks \citep[e.g.,][]{bouwman2008}, including the disk of $\beta$ Pic itself \citep{lu2022}. Such studies provide constraints on the average composition of dust in the disk, which is a mixture of dust originating from many different bodies. Through spectroscopic studies of exocomets, dust from an individual body can be studied, providing for a potentially very clean analysis of its composition. Here, we have performed a test case to evaluate to which extent an exocomet exhibiting silicate emission features might be spectroscopically characterized. 

For the purpose of this test, we assume a  comet with the same effective coma size as in Sect. \ref{s:orbit} (5~$R_{\rm E}$) at an instantaneous star-comet separation of 0.6 au. We then model the spectral output of such a comet, assuming that it exhibits silicate emission, based on different compositions of the silicates. Following \citet{lu2022}, we source models of Enstatite from \citet{chihara2002}, Forsterite from \citet{zeidler2015}, and Olivine and Pyroxene from \citet{jaeger1994} and \citet{dorschner1995}. We sequentially model one species at a time and note its impact on the output spectrum. This is obviously a highly idealized procedure, since a comet may in reality contain a mix of different silicate species; but the point of the analysis is to probe whether mineralogical variations can be confidently measurable in the first place, rather than attempt to realistically model the actual composition of an exocomet, which we have no way of knowing prior to any actual measurements. 

The spectrum of the hypothesized exocomet is a combination of blackbody emission and silicate emission, so we start from a blackbody spectrum in the same way as for the featureless comets in Sect. \ref{s:frequency}. We then redistribute half of this flux into silicate emission -- in other words, we halve the blackbody flux, normalize the pure silicate emission spectrum (of an individual species) to the same total flux (within the \textit{LIFE} spectral range), and add the two together. The equal flux redistribution is arbitrary, but it represents a case which has less fractional silicate emission than, e.g., Hale-Bopp, but more fractional silicate emission than a featureless spectrum like that of Halley's comet. It is therefore a mix encompassed by objects known to exist in nature. We then use the composite models in LIFEsim, using the same observational parameters as in the preceding sections. In addition to the standard 10 h simulated observations as before, we also ran 100 h observations to illustrate the effect of increased observing time on the achievable $S/N$.

\begin{figure*}[htb]
\centering
\includegraphics[width=18cm]{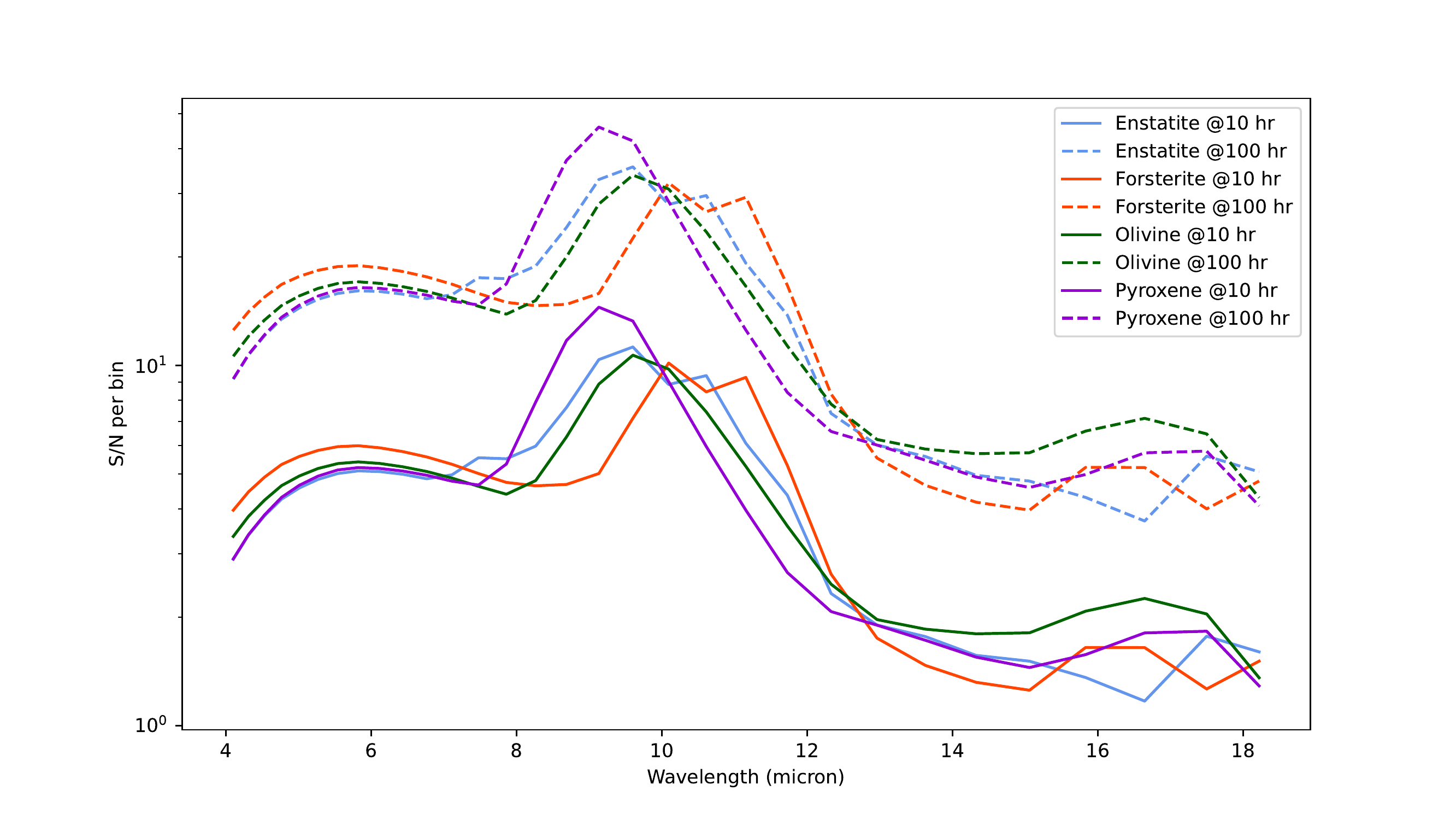}
\caption{Simulated $S/N$ values per wavelength bin using LIFEsim, for four cases of possible cometary emission with different silicate species: Enstatite (blue), Forsterite (red), Olivine (green) and Pyroxene (purple). Solid lines denote 10 h observations, while dashed lines represent 100 h observations.}
\label{f:snrspectrum}
\end{figure*}

\begin{figure}[htb]
\centering
\includegraphics[width=9cm]{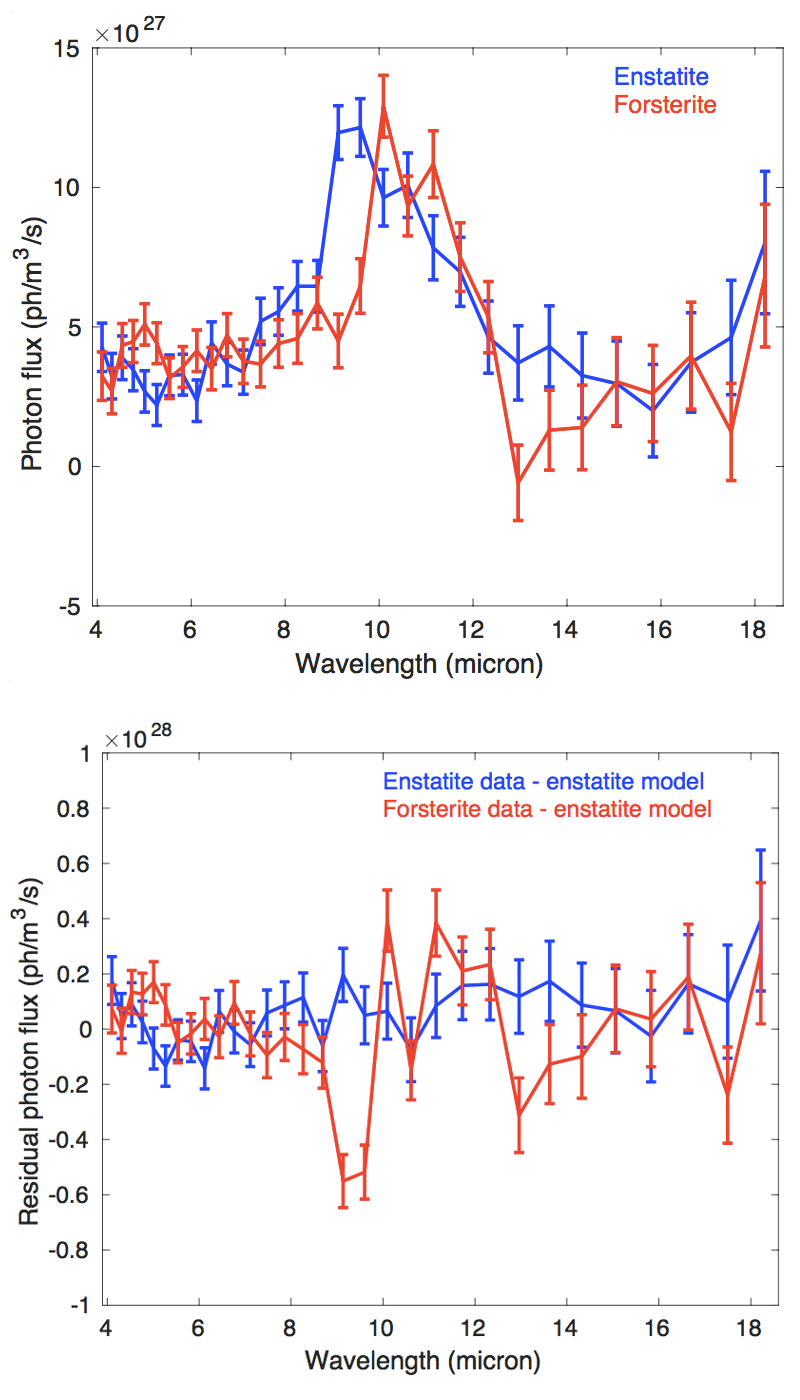}
\caption{Example of exocomet spectroscopy with \textit{LIFE}. Top: Simulated \textit{LIFE} spectra for two hypothetical exocomet cases, with noise levels set based on LIFEsim calculations for 10 h observations. The two cases correspond to pure Enstatite (blue) and pure Forsterite (red) compositions, respectively. The clearest distinction between the two cases is the peaking at shorter wavelengths for the Enstatite case than in the Forsterite case. Bottom: Residuals of the same two spectra after subtraction of a pure Enstatite model, leaving significantly smaller residuals in the true Enstatite case than the Forsterite case, relative to the error bars.}
\label{f:simspec}
\end{figure}

The output $S/N$ as function of wavelength for the simulations of each separate silicate species is shown in Fig. \ref{f:snrspectrum}. In order to more clearly show how these $S/N$ values affect the measured spectra, we selected Enstatite and Forsterite as two example cases and added random Gaussian noise to the respective spectral models with noise levels based on the simulated 10 h $S/N$ values at each sampled wavelength. The results are shown in Fig. \ref{f:simspec}. It can be seen that at least for this relatively large comet, a single 10 h observation is sufficient to start distinguishing pure Enstatite from pure Forsterite, with e.g. Forsterite being distinguishable from an Enstatite model by more than 5$\sigma$ in some spectral channels. Hence, some basic mineralogy can be performed in this circumstance, and distinguishing the presence of silicate emission from pure blackbody emission is easier still.

\section{Discussion}
\label{s:discussion}

As we saw in Sect. \ref{s:frequency} and Sect. \ref{s:crowding}, large comets may potentially be so abundant in the $\beta$ Pic system that exocomet observations with \textit{LIFE} could become confusion limited -- i.e., each resolution element close to the central star may contain several comets simultaneously. Such a scenario would essentially yield a diffuse cloud around $\beta$~Pic, similar to an exozodiacal disk. Since the beam combination nulling method used for detecting point sources with \textit{LIFE} cancels out symmetric features, the majority of this cloud would simply not get picked up by the observations. Since the expected brightness distribution of the exocomets is highly skewed, with a large number of small comets and a small number of large ones, the brightness distribution observed by \textit{LIFE} could generally be expected to be dominated by a small number of discrete point sources, corresponding to the brightness peak of the exocomet distribution. Since the flux distribution observed by \textit{LIFE} has to be synthesized from a limited number of interferometric baselines, mapping such a distribution might still be a challenge. Future work will be required to evaluate to which extent a large simultaneous number of point sources can be distinguished and mapped out by \textit{LIFE}. 

The above discussion is based on a scenario in which the $\beta$~Pic cometary distribution is relatively uniformly spread in terms of orbital inclinations ($\phi = 0.01$). If the distribution is much more concentrated closely to the line of sight $\phi \sim 1$, overcrowding may be much less of an issue. In either case, the high implied rate of bright comets around $\beta$~Pic clearly indicates that exocomet detection is a feasible task, at least in young systems with a high dynamical activity. If the $\beta$~Pic system is in fact over-crowded with bright exocomets, this speaks in favor of exocomet detection in other, slightly less active systems. In this context, $\epsilon$ Eri and Fomalhaut may be particularly promising targets, for several reasons: Both have bright debris disks but are an order of magnitude older than $\beta$~Pic, which means that their cometary activity is probably intermediate between $\beta$~Pic and regular main-sequence stars. Thus, they may feature a favorable balance between under- and overcrowding of exocomets in the field of view. Furthermore, both are at much smaller distances from us than $\beta$~Pic, such that the achievable $S/N$ (for equal-size comets) is much higher, and the inner solar system is much more finely resolved. Hence, while $\beta$~Pic is the template object for this study (simply due to the fact that it is the only system for which we have detailed information about the cometary distribution), it appears probable that a system like $\epsilon$~Eri is probably an ideal target for exocomet detection and characterization purposes.

The detectability of exocomets have two separate but similarly important implications: On one hand, it opens up the possibility to study a new class of astrophysical objects directly, potentially allowing for new insights into the building blocks of planetary systems. On the other hand, the overlap in observational properties between exocomets and terrestrial exoplanets means that we also need to consider that they may potentially be mistaken for each other. For example, a survey primarily directed toward habitable exoplanet detections may need to consider exocomets as one possible source of false positive signal. This should generally not be an issue for old single stars like the Sun, since observable exocomet events are very rare in such cases, as we noted in Sect. \ref{s:other}. However, in bright debris disk systems in particular, exocometary interloper events could be quite common, as we have seen, so among such targets it is an important consideration. It may also be relevant for certain types of multiple systems. 

For example, it has been hypothesized that the orbit of Proxima Cen around $\alpha$ Cen A and B could cause an enhanced cometary activity in the system, at least during certain epochs \citep{wertheimer2006}. In this context, we note that a particularly deep infrared imaging survey has been conducted around $\alpha$~Cen A and B with VLT/VISIR \citep{kasper2019}. The proximity of the system and depth of the observations allow for sensitivity to much smaller and more temperate planets than would otherwise be reachable with a ground-based 8m-class facility. A possible candidate for such a planet in the system has been proposed in \citet{wagner2021}, in the form of an extended feature close to the coronagraphic edge, which could possibly be interpreted as the trace of a planet undergoing orbital motion during the 33 days over which the observations spanned, at some point disappearing behind the coronagraph edge. The planet would be temperate (semi-major axis of $\sim$1.1 au), and have a size in the range of $\sim$3.3--7~$R_{\rm E}$. However, these ranges also overlap observationally with the exocomet population around $\beta$~Pic. Hence, if any similar population exists in the $\alpha$~Cen system, even if much less numerous than the $\beta$~Pic population, the observations could be equally well explained by a large exocomet. In this case, we do not consider either the planet or the comet scenario to be probable, since the reported candidate in \citet{wagner2021} is explicitly tentative, and could also be explained simply as a residual noise feature. Nonetheless, the example highlights that we are currently at a phase in high-contrast imaging and interferometry where we need to start seriously consider the observability of exocomets, both as scientific targets and as potential contaminants for other science cases.

The two most common classes of astrophysical false positives for directly imaged exoplanets are chance alignments of background stars, and unresolved background galaxies. In both of these cases, the most common method to distinguish between true and false positives is to observe the target system at least twice, with a time baseline of months or years between the epochs \citep[e.g.,][]{janson2011}. If the candidate point source does not share a common proper motion with the target star, it can be established as a background contaminant. For exocomets, this method could not be used in quite the same way for distinguishing them from planets, since the exocomet is bound to the star and therefore does share a common proper motion with it, just like a planet. However, as we saw in Sect. \ref{s:orbit}, astrometry at more than one epoch can provide a lower limit on the orbital eccentricity. While planets can have comet-like eccentricities (e.g. $>$0.7), classically habitable planets cannot, since they would spend a large fraction of their orbits outside of the habitable zone. It is therefore possible to at least distinguish habitable planets from exocomets in this manner. Furthermore, if followed up on a timescale longer than a few months, the exocomet would exit its near-periastron phase and turn quiescent, thereby disappearing entirely from view. A habitable planet can also temporarily disappear from view if on a highly inclined orbit, but would return back into view again over timescales of a few months, unlike a comet.

If an exocomet exhibits strong silicate emission (see Sect. \ref{s:spectrum}), then it can be very concretely distinguished from planets, which would not show such features. However, many Solar system comets do not exhibit clear silicate emission, but rather show blackbody-like spectra. Much like in the astrometric case, in this scenario we would not necessarily be able to distinguish exocomets from planets in general, but we would be able to distinguish them from classically habitable planets, since they would have atmospheres and therefore detectable atmospheric features \citep[e.g.,][]{quanz2022}. Distinguishing silicate emission features in exocomets would open a novel line of inquiry into the composition and differentiation processes of rocky and icy planetesimals, complementary to the results from polluted white dwarf, which indicate a fraction of exo-planetesimals to be geophysically processed \citep{bonsor2020,bonsor2022}. Constraining the composition and internal processing of wide-orbit planetesimals and exocomets can thus give a handle on late accretion phases and bombardment epochs of rocky exoplanets, altering their secondary atmospheres \citep{lichtenberg2021,lichtenberg2022}.

From a probabilistic viewpoint, we do not expect exocomets as potential false positives for terrestrial planets to be a big problem for studies around mature Sun-like or low-mass stars. However, infrared emission from magma ocean atmospheres in the aftermath of giant impacts among rocky exoplanets in young systems promise to reveal insights into the climate state of young, Hadean-like exoplanets \citep{lupu2014}, with distinct emission features in the optical \citep{boukrouche2021} and infrared \citep{bonati2019}. Therefore, it is an issue that should be kept in mind when planning observational strategies for a mission such as \textit{LIFE}. 

While the focus of this study is infrared detection and characterization of exocomets, particularly with \textit{LIFE}, there are of course also prospects for detecting exocomets through scattered light in the visible-light regime, if a sufficient contrast and sensitivity can be reached. As we have seen, comets of the type seen in transit around $\beta$ Pic have similar effective optical areas as rocky planets, and based on Solar system comets, their albedos are somewhat lower than, but broadly similar to, Earth-like planets. Thus, much like for \textit{LIFE} in the infrared, a mission scoped to image Earth-like planets at visible wavelengths \citep[such as LUVOIR, see e.g.][]{luvoir2019}, would probably have good prospects for being able to also image exocomets. Due to the potentially low comet albedo and fewer distinctive potential spectral features than in the infrared, visible light detection or characterization of exocomets might be more challenging than in the infrared case for a mission scoped primarily towards Earth-like exoplanets, but the yield may nonetheless be substantial in suitable debris disk systems. Any detailed discussion on the potential exocometary yields of a LUVOIR-like mission is beyond the scope of this study, but we note that if exocomets could be discovered at both visible and infrared wavelengths, the broader wavelength coverage would further enhance the information that can be extracted about their characteristics.

\section{Conclusions}
\label{s:summary}

As high-contrast direct imaging of exoplanets steadily improves with enhanced capabilities and ambitious plans for future missions, it is becoming increasingly important to consider the observability of exocomets in a direct imaging context -- both as a scientific opportunity, and as a possible source of planetary false positives. In this study, we have examined several aspects of exocomet direct imaging and spectroscopy with \textit{LIFE}. We have found that in systems with extreme cometary activities, such as $\beta$~Pic, there may be as much as hundreds of detectable exocomets at any given time, which would render the observations confusion-limited, although the resulting signal would be largely dominated by the few brightest comets in the distribution. Somewhat older, more nearby, and slightly less active debris disk systems such as $\epsilon$ Eri or Fomalhaut may be the ideal targets for maximizing actual exocomet yield and $S/N$.

Orbital characterization of detected exocomets is difficult due to the small fraction of orbital arc that is covered during its active phase at periastron passage, but setting a lower limit on the eccentricity is feasible, which is useful, not least for avoiding false identification of detected point sources in the target system. Some exocomets might also be expected to exhibit strong silicate emission features, which would be detectable within reasonable amounts of integration time, and could allow for the first mineralogical studies of individual exocometary bodies.

\begin{acknowledgements}
M.J. gratefully acknowledges funding from the Knut and Alice Wallenberg Foundation. The authors thank Laetitia Rodet for useful discussions and helping to verify the orbital analysis. This work has made use of data from the European Space Agency (ESA) mission {\it Gaia} (\url{https://www.cosmos.esa.int/gaia}), processed by the {\it Gaia} Data Processing and Analysis Consortium (DPAC, \url{https://www.cosmos.esa.int/web/gaia/dpac/consortium}). Funding for the DPAC has been provided by national institutions, in particular the institutions participating in the {\it Gaia} Multilateral Agreement. This work has made use of the Small Body Database managed by the Solar System Dynamics group at the Jet Propulsion Laboratory: \url{https://ssd.jpl.nasa.gov/}. The study has also made use of the CDS, NASA ADS, and NASA exoplanet archive services, as well as packages matplotlib \citep{hunter2007} and numpy \citep{harris2020}.
\end{acknowledgements}

\end{document}